\documentclass[%
 reprint,
 amsmath,amssymb,
 aps,
]{revtex4-1}

\usepackage{graphicx}
\usepackage{dcolumn}
\usepackage{bm}
\usepackage{color}

\begin{document}

\preprint{APS/123-QED}

\title{Density regulation in strictly metric-free swarms}

\author{D. J. G. Pearce${}^{1,2}$}

\author{M. S. Turner${}^{1,3}$}
\affiliation{University of Warwick Department of Physics${}^{1}$, Department of Chemistry${}^{2}$ and Centre for Complexity Science${}^{3}$, Coventry, CV4 7AL, United Kingdom}

\date{\today}

\begin{abstract}
There is now experimental evidence that nearest-neighbour interactions in flocks of birds are {\it metric free}, i.e. they have no characteristic interaction length scale. However, models that involve interactions between neighbours that are assigned topologically are naturally invariant under spatial expansion, supporting a continuous reduction in density towards zero, unless additional cohesive interactions are introduced or the density is artificially controlled, e.g. via a finite system size. We propose a solution that involves a metric-free motional bias on those individuals that are topologically identified to be on an edge of the swarm. This model has only two primary control parameters, one controlling the relative strength of stochastic noise to the degree of co-alignment and another controlling the degree of the motional bias for those on the edge, relative to the tendency to co-align. We find a novel power-law scaling of the real-space density with the number of individuals $N$ as well as a familiar order-to-disorder transition.
   

\end{abstract}

\pacs{Valid PACS appear here}
\maketitle



Swarming is the collective motion of a coherent group of animals. It appears in species of insects \cite{locustrings}, fish \cite{vortexshoalimage}, birds \cite{Murm} and mammals \cite{Bats}, including humans \cite{pedestrian}. In spite of some recent advances the actual mechanism for this collective motion in animal systems remains surprisingly poorly understood.\\
\indent Self-propelled-particle (SPP) models have historically been used as simple models that mimic this behaviour \cite{giardinarev}. A model due to Vicsek and co-workers \cite{vicsek} involves $N$ point particles moving with constant speed within a periodic box, which therefore has fixed density. Each particle updates its velocity in the next time step by first taking a local average of the velocities of neighbouring particles that are closer than a pre-defined interaction range, and then combining this with a noise term. In the absence of noise the particles within the system align perfectly with each other and a long range global order in the alignment is established, measured by the average centre of mass speed of the swarm $P=\vert\frac1N\sum_{i=1}^N{\bf v}_i\vert$. Conversely, in the limit of strong noise, particles perform uncorrelated random walks and there is no long-range order. On increasing noise there is known to be a transition between the ordered and disordered phase \cite{vicsek}. {The nature of this, and similar, transitions has been studied in depth \cite{tonertupre,tonertuprl,gregoire2004onset,chatepre,vicsek,Albano,baglietto2012criticality,nagy2007new,chate2008collective,baglietto2009nature} and recent studies have shown that the phase transition of the standard Vicsek model is continuous when the velocity is small \cite{Albano,baglietto2012criticality,nagy2007new}; and it has been argued that large velocities can lead to artefacts related to the boundary conditions \cite{baglietto2009nature}.
The nature of the transition has also been shown to depend intimately on the type of noise added to the particles \cite{aldana2007phase}, which can either be ``angular'', where a random angle is added to the orientation of the velocity of a particle, or ``vectorial'', in which a random vector is added to the velocity of a particle. In what follows we restrict our attention to ``vectorial'' noise, for which the nature of the order-disorder transition in the classic Vicsek model has been established to be discontinuous \cite{gregoire2004onset,baglietto2014complex}.}\\
\indent Recent experiments on flocks of starlings \cite{Ballerinistudy,ScaleFree} and human 
crowds \cite{metricfreepedestrians} have shown that the controlling interactions are more likely metric free, at least in large murmurations. This naturally constrains the underlying models to also have a metric-free character, ushering in a new class of metric-free models \cite{chatevoronoi,visualnetw,ballerinitopol}.
The topological Vicsek model \cite{chatevoronoi} differs from the original, metric-based, version in that neighbours are assigned to be those that form the first shell in a Voronoi tessellation, equivalent to being connected by edges under a Delauny triangulation \cite{delaunaybook}. This procedure has the advantage that it is completely metric free. It also has the satisfying feature that it has one less control parameter, lacking an interaction radius or number of nearest neighbours. When the resulting collective motion is studied in a periodic box (with vectorial noise) the system is found to support both a high density, ordered phase of aligned particles and a low density, disordered phase \cite{chatevoronoi}. {As the system size is increased (at constant density) the phase transition has been shown to take on a continuous character \cite{chatevoronoi}. }
Such a metric free model gives rise to an ordered swarm but the question of how it remains cohesive is still open \cite{Ballerinistudy}. Most studies involve either the introduction of attraction and repulsion terms \cite{chateAR,couzinAR,ARmodelreview}, confine the whole swarm to a periodic box \cite{ballerinitopol,chatevoronoi} or involve some kind of potential field \cite{vicsekCOM}. While these approaches are able to generate swarm cohesion they also introduce a metric to the system, either in the form of interaction ranges or a (periodic) system size. These often result in a characteristic density, at least implicitly, and lead to a density that is effectively constant and independent of $N$, which is not  seen in the data \cite{Ballerinistudy}.\\
\indent In this letter we propose a fully 3D, strictly metric free (SMF) model that also controls the density of a system of self-propelled particles in unbounded space. The topological Vicsek model, in which particles that move with some constant speed co-align with neighbours in their first Voronoi shell, supports global alignment order, regardless of swarm density.
To this structure we introduce a metric free surface term which, in behavioural terms, would correspond to individuals on the swarm exterior behaving differently. We first identify all individuals that could make a topological connection to a hypothetical individual separated to infinity (in any direction). These individuals lie on the convex hull of the swarm \cite{delaunaybook}. We refer to these as being on the {\it edge}, while all others are referred to as being in the {\it bulk}. We next introduce a metric-free inward motional bias for those on the edge. This is somewhat reminiscent of the effect of a surface tension in a thermodynamic system, an analogy that we will explore later. For those on the edge we define ``inwards" as the average of the vectors pointing to the Voronoi neighbors to that individual that are also on the convex hull. 
While other choices are possible, including the average of vectors to {\it all} neighbours, we believe that our choice is simple and has the property that the inward motional bias vanishes for individuals on a surface that has no convexity. This is in qualitative agreement with the existence of large swarms with relatively stable edges and also mimics the physics of a surface under tension. It is feasible for an individual on the edge of the swarm to identify others on the edge by first projecting the relative positions of all others (onto its retina), mapping them to points on the surface of a unit sphere. Those individuals on the edge of the flock can then easily be identified as lying on the boundary of the portion of the sphere containing these points{, indeed edge detection is known to be performed in the neural hardware of the visual cortex in higher animals \cite{birdbrainbook}}. Since a particle on the edge of the swarm changes its behaviour in order to rejoin the bulk, it also promotes mixing within the swarm and reduces the rearrangement time of the voronoi mesh.
{Because this model was motivated by the observed metric free nature of starling flocks \cite{Ballerinistudy,ScaleFree}, we restrict ourselves to motion in 3D so as to best compare with large murmurations of birds such as starlings. Different organisms have been observed to follow very different schemes \cite{guttal2012cannibalism,portugal2014upwash}. We propose a minimal model to address the question of metric free density control, but the model could easily be expanded to capture many more features of bird flocks by the inclusion of additional refinements such as blind angles, flight physics etc.}
\\
\indent The SMF model has a number of features that are appealing in behavioural terms. First, its metric free character means that an animal isn't required to accurately judge and compare distances. It only requires that an individual recognise when it is on the {\it edge} and identify its neighbours with the most extremal (angular) positions within its view.
Secondly, the position of neighbours on the edge give each of them information on the boundary, and hence the shape of the swarm as a whole; information that is not available from local interactions in the bulk. 
Finally, assigning control of density to those that reside on the edge also means that it lies with those that are most exposed to predation. It seems reasonable that they should be those that select the swarm structure. Those in the bulk are the safest from predation, so are least likely to try to change their relative position or the properties of the swarm as a whole \cite{selfish}. 
Recent studies on large starling flocks have also noted that there is a local rise in bird density toward the edge of flocks, which would tend to support some distinction between edge and bulk, such as we introduce here \cite{Ballerinistudy,Cavagnaboundaryinfo}.
\\
\indent Our SMF model is defined as follows. At discrete time $t$, particle $i$ has position $\underline{r}_i^t$ and direction of motion $\hat{\underline{v}}^{t}_{i}$, with $\hat{\underline{r}}_{i,j}^{t}$ representing the unit vector pointing from particle $i$ toward particle $j$ at time $t$. These are updated every time step according to equations Eq.~(\ref{eq:r} - \ref{eq:f}). The neighbors, $B_i$, are those forming the first shell around particle $i$ in a Voronoi tessellation constructed from the particle positions at time $t$. If particle $i$ is in the set $C$ of all those on the convex hull of the system, the neighbors, $S_i=B_i\cap C$ are those that also lie on the surface. 

The model is controlled by two parameters. 
First $\phi_e$, the strength of the ``edge'' effect, is the relative weight of alignment to the inward movement bias for particles on the edge.
The second parameter is  the strength of the noise, $\phi_n$. See Eq.~(\ref{eq:r} - \ref{eq:f}). 

\begin{eqnarray}
\underline{r}^{t+1}_{i}= \underline{r}^{t}_{i}+v_{0}\hat{\underline{v}}^{t}_{i}
\label{eq:r}
\\
\underline{v}^{t+1}_{i}= (1-\phi_{n})\hat{\underline{\mu}}_{i}^{t}+\phi_{n}\hat{\underline{ \eta }}^{t}_{i}
\label{eq:v}
\\
\underline{\mu}_{i}^{t}=(1-f_{i})\frac{ \langle\hat{\underline{v}}^{t}_{j}\rangle_{j\in B_i }}{| \langle\hat{\underline{v}}^{t}_{j} \rangle_{j\in B_i } |}+f_{i} \langle\hat{\underline{r}}^{t}_{ij} \rangle_{j\in S_i} 
\label{eq:mu}
\\
f_{i} =\begin{cases}\phi_{e} & \underline{r}^{t}_{i} \in C\\0 & {\rm otherwise} \end{cases} 
\label{eq:f}
\end{eqnarray}
where $\hat{\underline{ \eta }}$ denotes a random unit vector and $v_0$ is the speed with which the particles move; throughout a hat   $\hat{}$   denotes a normalized unit vector and angular brackets $\langle\cdots \rangle$ denote an average over the subset indicated. {We structure our model so that every individual experiences a similarly weighted noise contribution, controlled by $\phi_n$ in Eq.~\ref{eq:v}. According to Eq.~\ref{eq:mu}  each individual first decides a (deterministic) direction before any noise is introduced. This direction is determined by either a) simply co-aligning with neighbours - in the case where the $i^{\rm th}$ individual is not on the surface, $f_i = 0$ according to Eq.~\ref{eq:f}, or b) resolving a  linear combination of the co-alignment direction and the one that arises from the metric free surface term, these being weighted by factors $1-\phi_e$ and $\phi_e$ respectively - when $i$ is on the surface, $f_i = \phi_e$ according to the same Eq.~\ref{eq:f}. This means that the combination of a deterministic ``preferred" direction and a random noise vector is similar for each individual, irrespective of location.}

\begin{figure}[h]
\includegraphics[width = \columnwidth]{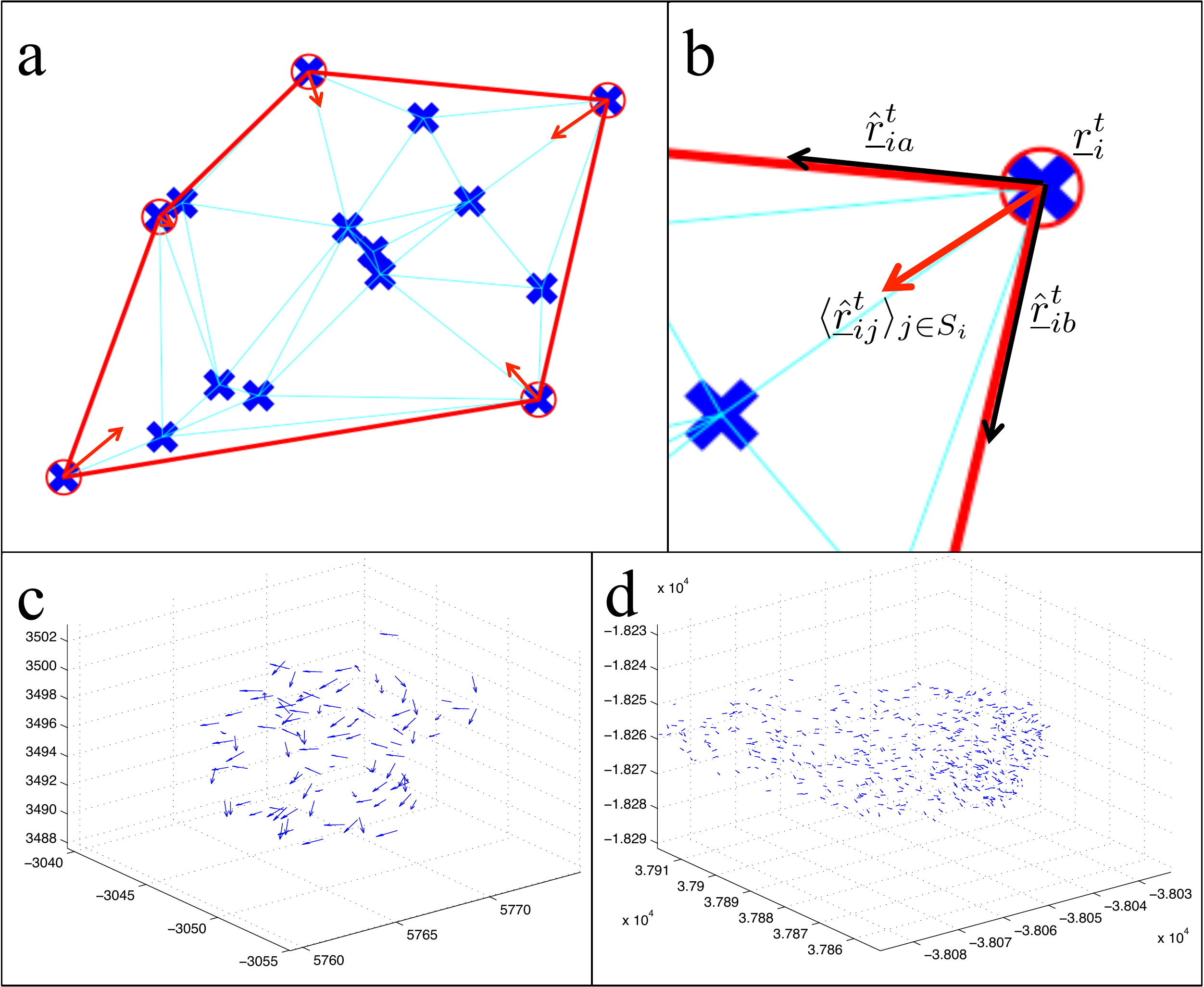}
\caption{\label{fig:vorsnap} {(a) An sketch showing of the topological constructions included in Eq.~\ref{eq:mu}. Here an arrangement of points ($\underline{r}^{t}_{i}$) are shown as crosses in the 2D plane, for simplicity (the model is fully 3D), with those lying on the convex hull circled in red (for which $f_i \not= 0$ in Eq.~\ref{eq:f}). The Delaunay triangulation is shown in cyan, and the red lines denote the subset which connects points on the convex hull. Therefore, points connected by a cyan (or red) line are those in $B_i$, hence will co-align, and points connected by a red line are those in $S_i=B_i\cap C$, hence contribute the the metric free surface term (displayed as a red arrow for each point on the convex hull.) b) Detailed sketch showing how the metric free surface term appearing in Eq.~\ref{eq:mu} is constructed for a general particle $i$. This depends on the vector $\langle\hat{\underline{r}} ^{t}_{ij} \rangle_{j \in S_i}$ (red arrow) which is the average of unit vectors to adjacent individuals {\it that are also on the convex hull}, $\underline{r} ^{t}_{ia}$ and $\underline{r} ^{t}_{ib}$ for particles $a$ and $b$, respectively (black arrows). 
c) and d) Snapshot of a realisations of the model for 100 and 500 particles, respectively. Both have a noise weighting of $\phi_n = 0.45$. Both flocks have a polarisation of $P \approx 0.75$.}}
\end{figure}

{Fig.~\ref{fig:vorsnap}a shows an example of the topological constructions included in Eq.~\ref{eq:mu}. The metric free surface term for each individual is taken as the average of the unit vectors pointing to adjacent points that are on the convex hull ($\langle\hat{\underline{r}} ^{t}_{ij} \rangle_{j \in S_i}$). This has a magnitude in [0,1] and is greatest when the angle between the $\underline{\hat{r}}_{ij}$ is small. Therefore individuals that are `exposed' outliers associated with the sharpest kinks in the convex hull, will have a stronger drive to rejoin the swarm. This construction is shown in more detail in Fig.~\ref{fig:vorsnap}b.}

\begin{figure}[h]
\includegraphics[width = \columnwidth]{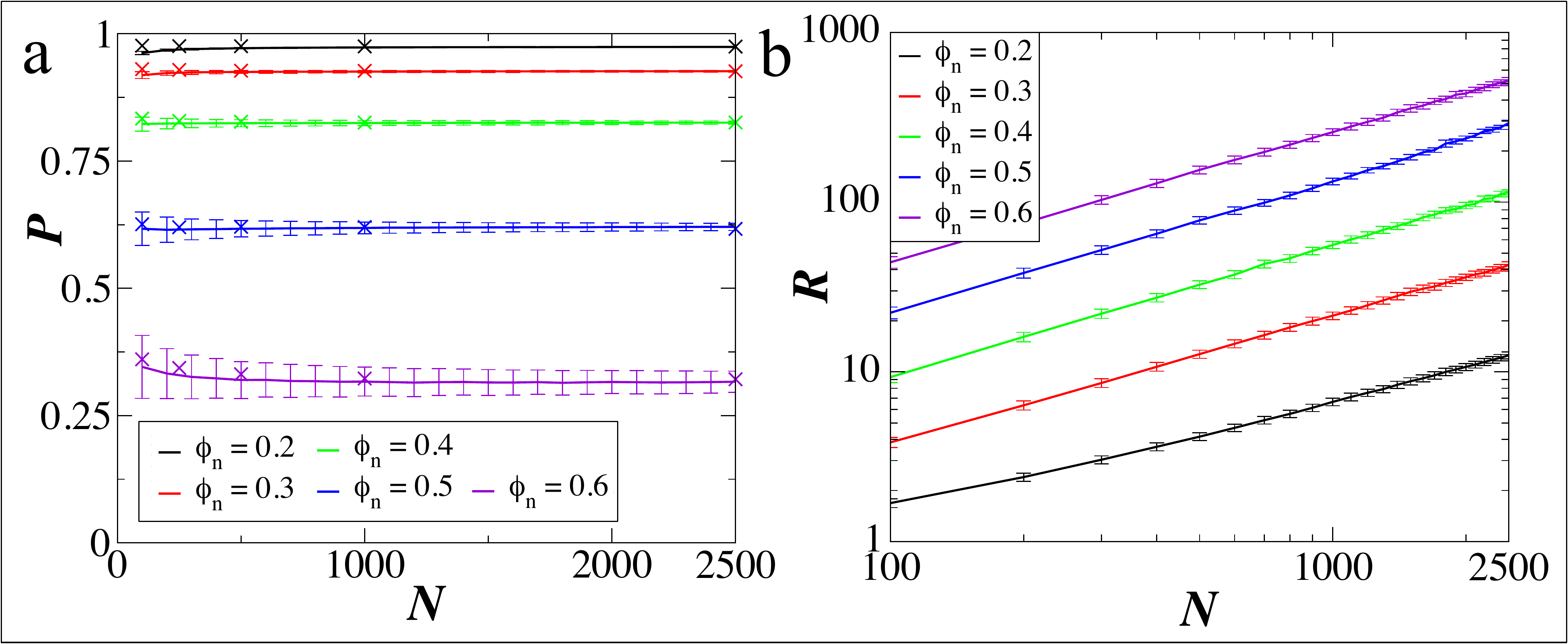}
\caption{\label{fig:nscale} (a) Global order, $P$, and (b) Swarm spatial extent, $R$, for swarms containing different numbers of individuals, $N$, and with varying levels of noise, $\phi_n$ (the key given in panel a also applies to b). Each point on the figures corresponds to an average ($\pm \sigma$, one standard deviation) over 40,000 simulation time steps following a 10,000 time step pre-equilibration period, significantly longer than either the density or order autocorrelation times.  $\phi_e = 0.5$ for all simulations except the points represented by the $\times$ in (a) corresponding to the values achieved for $\phi_e = 0$, i.e. a benchmark topological Vicsek model that disperses in space.}
\end{figure}

\begin{figure}[h]
\includegraphics[width = \columnwidth]{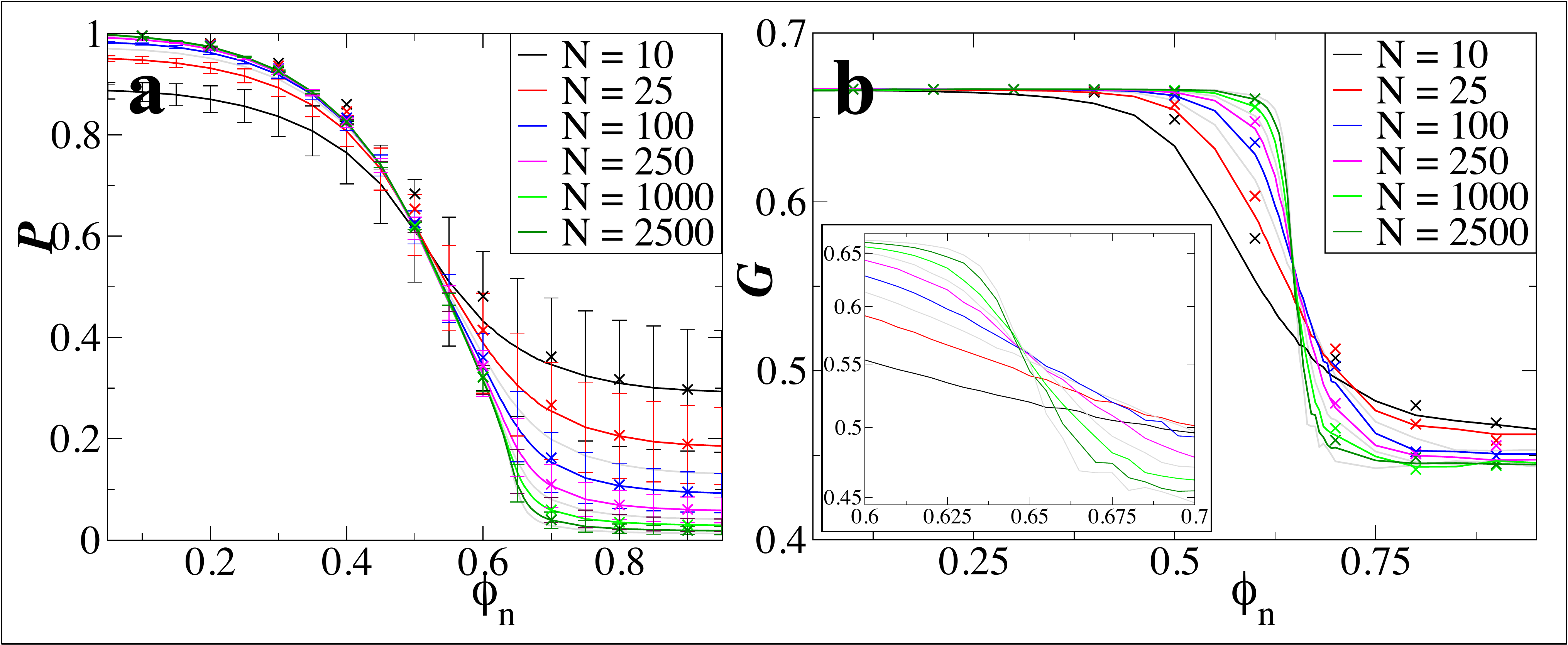}
\caption{\label{fig:trans}  (a) Global order, $P$, and (b) Binder cumulant $G$ \cite{BinderC} as a function of the noise strength $\phi_n$ and enlarged area near the transition with logarithmic scale (inset). There is a continuous transition from an ordered to a disordered state as the noise increases and this is independent of the number of individuals $N$. $\phi_e = 0.5$ for all simulations except the points represented by the $\times$ corresponding to the values achieved for $\phi_e = 0$, i.e. a benchmark topological Vicsek model that disperses in space. }
\end{figure}

This set of equations can be solved iteratively and gives rise to a coherent and ordered swarm in both two and three dimensions, see {Fig.~\ref{fig:vorsnap}c,d and }S.I. for movies. Because the equations are completely metric free, the choice of units is somewhat arbitrary{, see SI for details. The insensitivity of our model to the value of $v_0$ may simplify the task of constructing continuum models \cite{degond2008continuum,czirok2000collective}}. The only dimensional units are the distance travelled per time step, $v_{0}$, and the duration of a time step, both of which are set to unity, thereby defining our length and time units. In unbounded space the model with $\phi_e=0$, is found to support either an ordered or a disordered state depending on the noise level, $\phi_n$, with a continuous transition between the two. This is despite the fact that, without a term acting at the edge to bound the swarm it is undergoing continuous spatial expansion {and a corresponding decrease in density} \cite{chatevoronoi}. This result is independent of the number of particles for sufficiently large swarms, $N \gtrsim 500$. Without any terms to control the density the swarm disperses in time, approaching zero density. Although this is not a realistic model for swarms with spatial cohesion it does provide a benchmark global order (and order-disorder transition) for a swarm in the absence of modifications that suppress fracture and is shown as $\times$'s on Fig.~\ref{fig:nscale}a and Fig.~\ref{fig:trans}a,b (there is no such data on Fig.~\ref{fig:nscale}b since the continuous spatial expansion does not produce a non-zero steady state density). Here, and in what follows, we set $\phi_e = 0.5$ for simplicity, providing equal weighting to co-alignment and inward bias in Eq.\ref{eq:mu} for individuals on the edge of the swarm. The effects of varying $\phi_e$ itself are covered further in the SI. This leaves only one free control parameter, the noise strength $\phi_n$. The similarity between the benchmark properties of the models shown in Fig.~\ref{fig:nscale}-\ref{fig:trans}, both with and without the new surface term, are included merely as supporting results to show that the introduction of the surface term doesn't ``break" the other well known properties of the model.

Swarms generated by our SMF model can achieve highly ordered states. A high global order parameter $P$ emerges naturally when the noise is sufficiently low, see Fig.~\ref{fig:nscale}a. Note that the precise value of $P$ is roughly independent of the size of the number of individuals in the swarm, $N$. This implies that the swarm maintains a particular level of global order without the need for the individual members to comprehend (and respond to) the size of the swarm in which they reside. This could explain how relatively simple animals can participate in swarms which vary in size by several orders of magnitude without the swarms qualitatively changing their behaviour \cite{Ballerinistudy}.

The surface term also has very little effect on how the swarm forms an ordered state: for large $N$ the swarms reach roughly the same global order (see $\times$'s on Fig.~\ref{fig:nscale}a). This is consistent with the fact that the proportion of particles in $C$ (on the convex hull) decreases with $N$. 
In order to assess the spatial cohesion of the swarm we first define the swarm's spatial extent, $R$, as the average voronoi neighbour separation $R_d$ multiplied by the cube route of the number of individuals, i.e. $R = R_{d}N^{1/3}$. While alternative definitions are possible this measure has the attractive feature that it isn't strongly biased by a single outlying individual that may, by chance, have moved some distance away from the others.  Fig.~\ref{fig:nscale}b shows that the SMF model generates swarms with a well defined equilibrium extent, $R$, (and therefore density) which appears to follow a power law $R \sim N^{0.8}$. Hence the SMF model is able to support ordered swarms that remain spatially cohesive in unbounded 3D space; a robust feature of animal swarms. 

A transition between an ordered and disordered state occurs for swarms generated by the SMF model, Fig.~\ref{fig:trans}a. The nature of the transition converges for large $N$. The presence of a smooth transition is confirmed by the monotonic nature of the Binder cumulant, $G = 1-\frac{\langle P^{4} \rangle_{t}}{3\langle P^{2} \rangle^{2}_{t}}$ \cite{BinderC}, Fig.~\ref{fig:trans}b. This is very similar to the transition observed for the benchmark topological Vicsek model with $\phi_e=0$ that disperses in space (shown as $\times$'s), again confirming that the new surface term controlling the density doesn't compromise the swarms ability to form an ordered (or disordered) state. The continuous nature of the phase transition is also found to b independent of th the choice of $\phi_e$, see SI for details.



The fact that the location and continuous nature of the transition are independent of $N$ implies that a group of swarming animals can occupy the \textit{sweet spot} near the inflection point in the order-disorder transition without the need for any individuals to adjust their behaviour as the number of individuals in the swarm changes.  This would seem to be compatible with observations on swarming animals in the wild, which often occupy an intermediate state which displays high local order with a short de-correlation time; this is likely linked to evolutionary fitness, meaning a large group of animals can be both ordered and react quickly.

The metric free nature of interactions between starlings within large flocks has been reported by the STARFLAG collaboration \cite{Ballerinistudy}. It was found that each bird's velocity is highly correlated with that of a fixed number, $n_c$, of its nearest neighbours, regardless of the sparseness of the flock, $R_1$, defined as the average nearest neighbour separation. From this it was inferred that the orientational correlation between two individuals of a flock depends on topological rather than metric distance \cite{ballerinitopol}. This was confirmed by measuring a \textit{topological range}, $n_{c}^{1/3}$, which was directly observed to be constant for flocks of varying density. By contrast the metric range, defined as the average distance between birds with highly correlated velocities, scales linearly with the sparseness \cite{ballerinitopol,ScaleFree,Ballerinistudy}.

In the SMF model, we can define the metric correlation length scale, $R_d$, as the average distance to the Voronoi neighbors, and the topological correlation length scale as the cube root of the average number of Voronoi neighbours, $N_{d}^{1/3}$. These are analogous to the quantities calculated from measurements taken on flocks of starlings in the wild \cite{ballerinitopol}. Since the nearest neighbour separation, $R_1$, is not a parameter that is under direct control in the SMF model, we adjust it by changing the number of members, $N$, and noise level, $\phi_n$, of the swarm and measuring the resulting separation. 

\begin{figure}[h]
\includegraphics[width = \columnwidth]{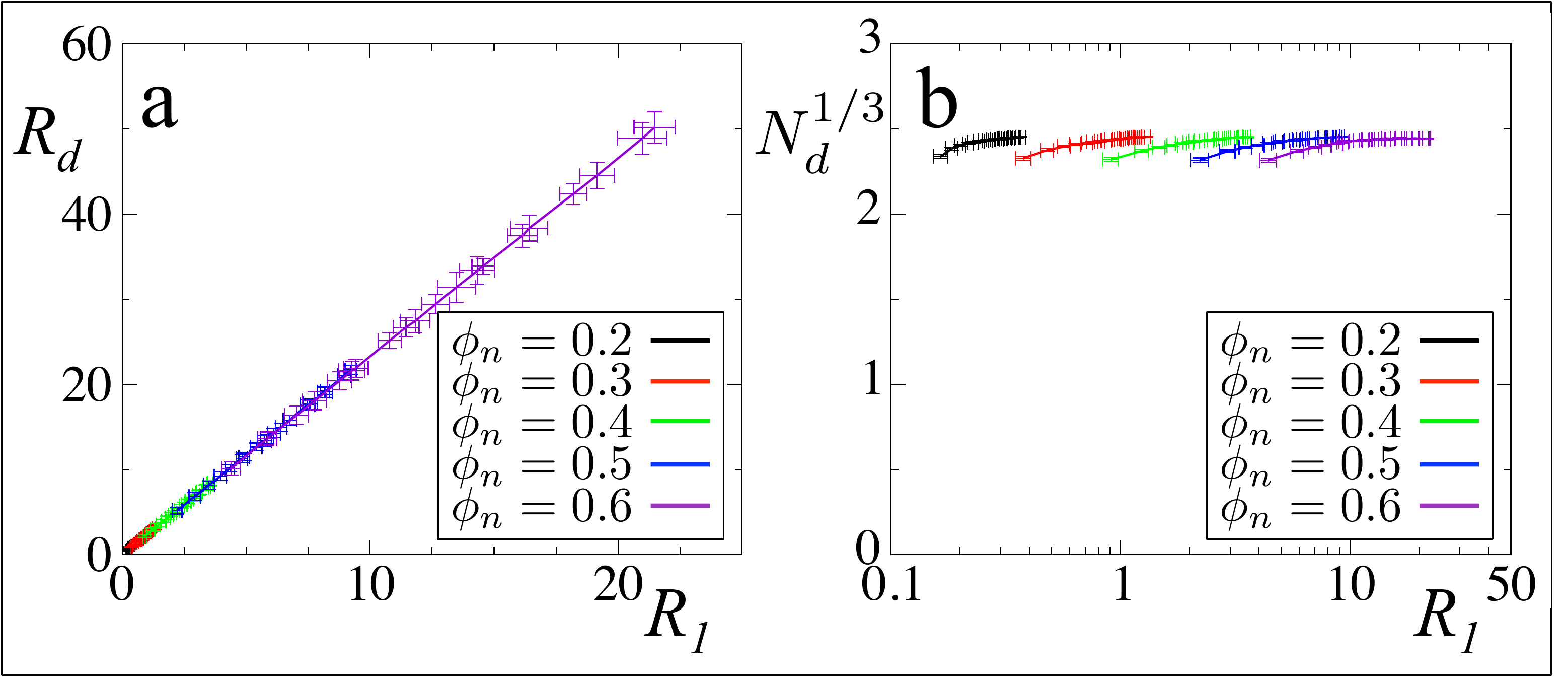}
\caption{\label{fig:corrlen} The variation of the correlation ranges, (a) metric, $R_d$, and (b) topological, $N_{d}^{1/3}$, as a function of the nearest neighbour distance $R_1$ for varying noise levels. The multiple data points correspond to different values of $N$. These trends closely agree with data on the metric and topological length scales observed in starling flocks \cite{ballerinitopol}. $\phi_e = 0.5$ for all simulations.}
\end{figure}

Fig.~\ref{fig:corrlen}a shows a linear relationship between the metric range, $R_d$, and sparseness, $R_1$. Hence $R_d$ scales with the size of the swarm, a consequence of the metric free nature of the model. Conversely, $N_{d}^{1/3}$ remains roughly constant as the sparseness of the swarm changes, confirming the fact that the model proposed here is indeed completely metric free, with purely topological interaction ranges. The observed upward trend for denser swarms is due to the fact that these are smaller, $N<500$; an amorphous swarm has a higher proportion of its members on the surface, which, in turn, have fewer Voronoi neighbours. This is in agreement with the analogous quantities for the topological and metric length scales measured in starling flocks \cite{ballerinitopol}.

The spatial extent $R$ of SMF swarms follows a power law relationship with the number of particles, $N$, (Fig.~\ref{fig:nscale}b). Since the co-alignment of particles may have non-linear effects on the swarm density (connected to the breaking of Galilean invariance by the convention that $v_0$ is constant) we simplify the SMF model to eliminate co-alignment, i.e. Eq.\ref{eq:mu} becomes
\begin{equation}
\underline{\mu}_{i}^{t+1}=f_{i} \langle\hat{\underline{r}} ^{t}_{ij} \rangle_{j\in S_i}\label{nocoalign}
\end{equation}
In this way we hope to gain a clearer understanding of  this power law relationship, first via a simple analogy with an ideal gas: In the absence of particle-particle co-alignment each member of the swarm resembles a gas molecule, only feeling an anisotropic force, on average, when it reaches the convex hull of the swarm; much like a bubble of ideal gas.

\begin{figure}[h]
\includegraphics[width = \columnwidth]{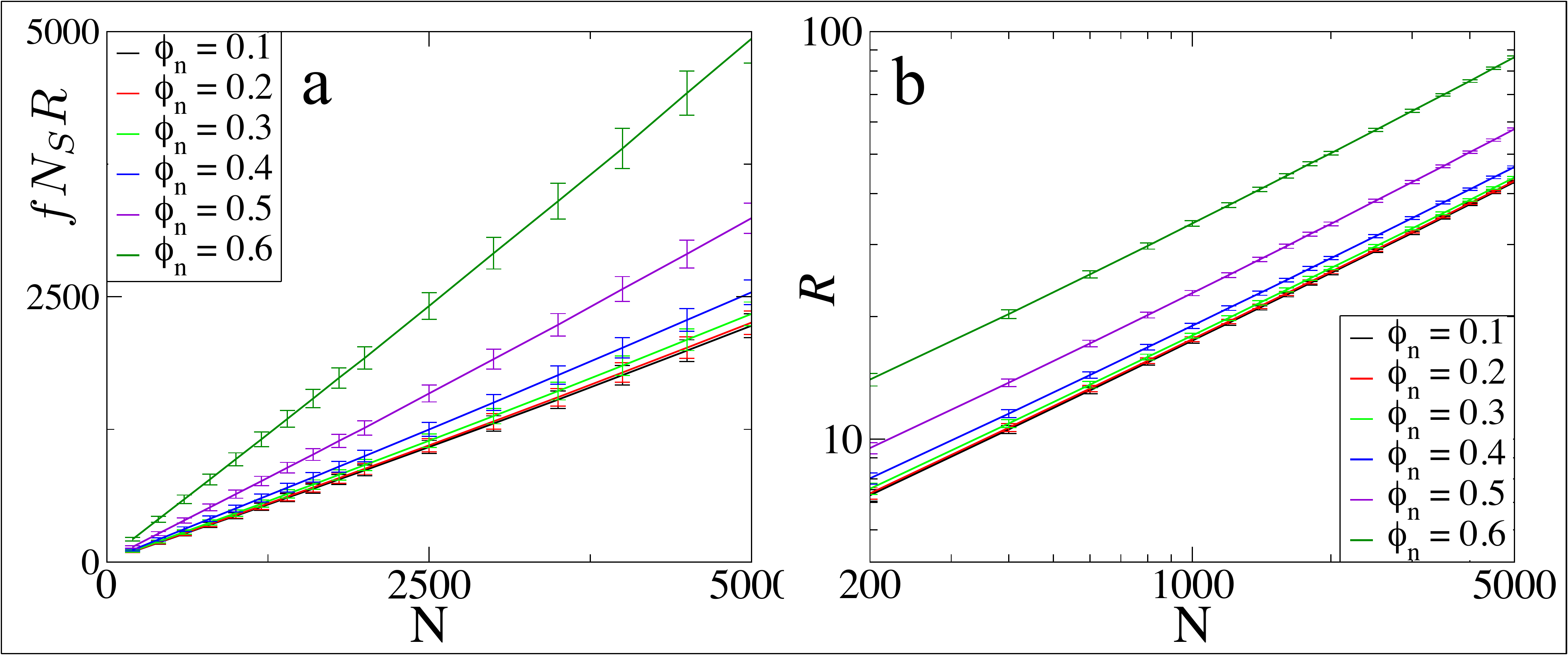}
\caption{\label{fig:gasscale} a) For swarms without co-alignment, the product of the number of particles on the surface of the swarm $N_S$, their inward motional bias ($\sim$ force) $f$ and the radius of the swarm $R$ is seen to scale linearly with the number of particles $N$. As discussed in the text this provides supporting evidence that this state that is analogous to an ideal gas. b) A power law relationship between $R$ and $N$ is observed. }
\end{figure}

Starting from the ideal gas law, $pV \sim NT$, and substituting in $p \sim F/A$ for the pressure we arrive at $FV/A \sim FR \sim NT$. Here N and R are naturally the number of individuals and the swarm radius, respectively. The analogue of the total inward force on the swarm, $F$, is proportional to the number of particles in $C$ (on the surface) $N_S$, and to the average inward motional bias, $f = \langle\langle\hat{\underline{r}}_{ij} \rangle_{j\in S_i} \rangle_{i\in C}$. Since the speed $v_0$ of the particles is fixed, the temperature, $T$, is assumed to rely only on noise, hence $T(\phi_n)$. Substituting back in for $F$ gives us the relation, $fN_{S}R \sim NT(\phi_n)$, confirmed for swarms of such gas-like particles, see Fig.~\ref{fig:gasscale}a. The power law relationship between $N_{S}$ and $f$ with $N$ leads us to expect a similar relationship between $R$ and $N$, which is confirmed in Fig.~\ref{fig:gasscale}b. The nature of this relationship is surprisingly robust to the introduction of co-alignment between particles, resulting in the power law observed in Fig.~\ref{fig:nscale}b.

In this letter we have shown how the primary features of a flock of starlings, including spatial cohesion, order, low autocorrelation times and metric-free correlation lengths can be generated using a fairly simple, strictly metric free (SMF) model. This model gives rise to a power law relationship between the spatial extent of the swarm and the number of individuals which arises as a result of the different role of individuals on the edge of the swarm.

As well as giving good agreement with observations of animal systems (despite its significant simplifications), this model has the appealing feature that it only requires  individuals to perform relatively simple measurements/computations: the relative position and velocity of a finite (and modest) number of nearest neighbours and an awareness of when an individual is itself on the surface of the swarm. These are cognitive tasks that would seem to be within the ability of a large number of swarming animals, in contrast to models, e.g. that regulate density using long-ranged (metric based) attraction. These have algorithms that involve $O(N)$ computations per individual, $O(N^2)$ overall, per timestep. This property of the SMF model may help to explain how animals with relatively limited abilities are able to organise themselves into impressive displays of coordinated behaviour. 

\begin{acknowledgments}
This work was partially supported by the UK Engineering and Physical Sciences Research Council through the MOAC Doctoral Training Centre (DJGP) and grant EP/E501311/1 (a Leadership Fellowship to MST). We also acknowledge Computational Geometry Algorithms Library (CGAL) which was used to create the simulations \cite{cgal}.
\end{acknowledgments}



\end{document}


\preprint{APS/123-QED}

\title{Density regulation in strictly metric-free swarms\\ {\rm Supplementary Information}}

\author{D. J. G. Pearce${}^{1,2}$}

\author{M. S. Turner${}^{1,3}$}
\affiliation{University of Warwick Department of Physics${}^{1}$, Department of Chemistry${}^{2}$ and Centre for Complexity Science${}^{3}$, Coventry, CV4 7AL, United Kingdom}

\date{\today}

\maketitle

\section{Speed insensitivity}

\begin{figure}[!h]
\includegraphics[width = \columnwidth]{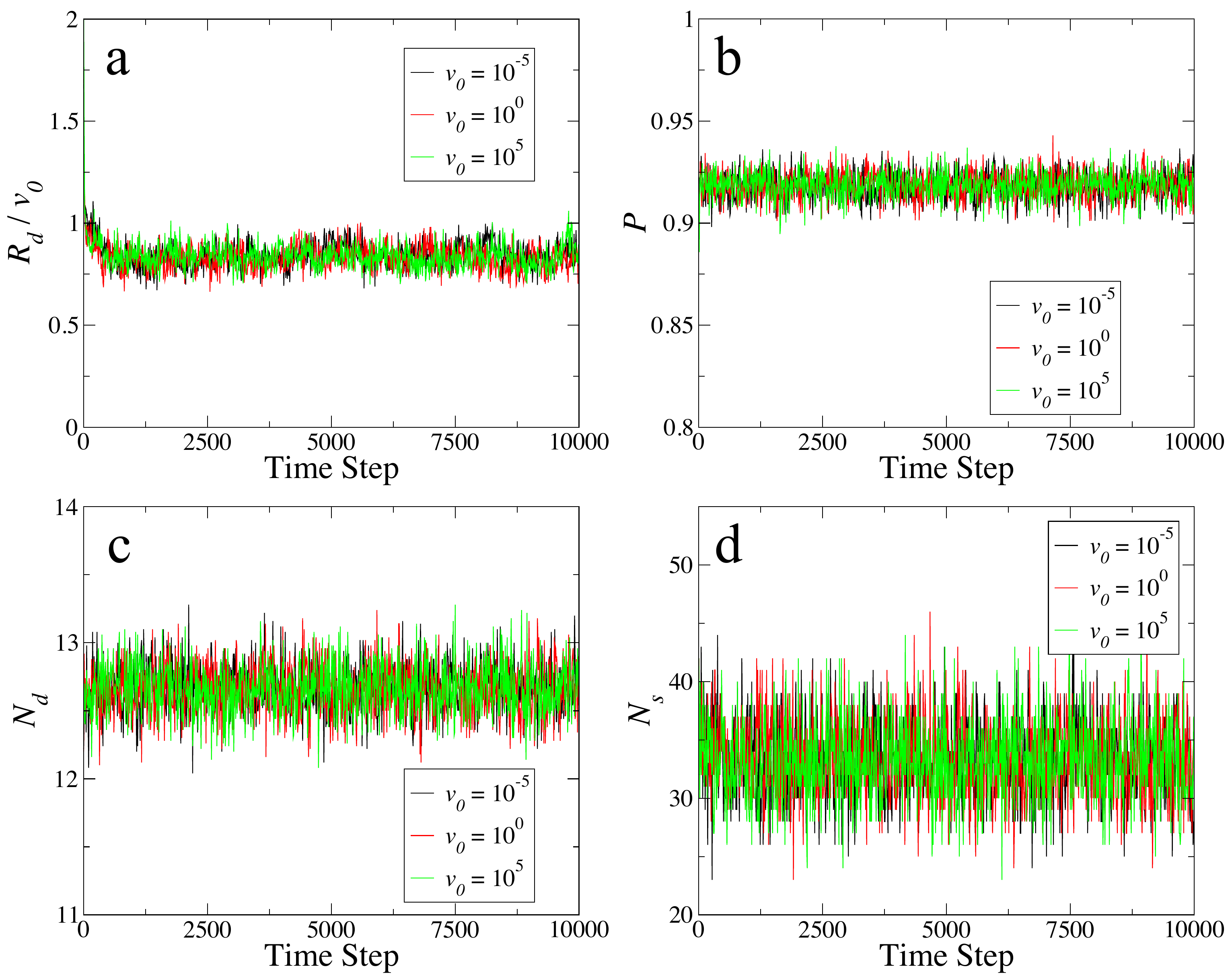}
\caption{\label{fig:v0} Confirmation of the arbitrary nature of the choice of $v_0$, due to the metric free nature of the model. Panel (a) shows the average distance to Delaunay neighbours, (b) shows the polarisation, (c) the average number of Delaunay neighbours, and (d) the average number of particles on the convex hull, all as a function of time for various choices of $v_0$, as shown. In every case all measurable quantities are unaffected by a change in $v_0$ across 10 orders of magnitude.}
\end{figure}

The model only features one dimensional unit, $v_0$, which defines the length moved by a particle within a time step. We set $v_0$ to unity, thereby defining our length and time units, but as mentioned in the main text this is somewhat arbitrary. However, the results are completely insensitive to the value of $v_0$ due to the metric free nature of the model itself. To confirm this we ran simulations with values of $v_0$ that vary by ten orders of magnitude and observed no difference in the behaviour of the model, see Fig.~\ref{fig:v0}.  The density of the swarm is invariant to the value of $v_0$ in the sense that even though doubling the speed may result in a doubling of the inter particle distance it also doubles the only unit in which it can be measured, Fig.~\ref{fig:v0}a. The polarisation is a function only of the direction of motion of each individual particle, so is speed-invariant, see Fig.~\ref{fig:v0}b. Because quantities like the number of Voronoi neighbours and the number of particles on the convex hull are purely topological constructions they are invariant to such rescaling, see Fig.~\ref{fig:v0}c\&d. It is worth mentioning that if the model were to be modified to include some kind of inertial effects this could introduce another dimensional unit, changing these conclusions somewhat.

\section{Statistics at the edge}

\begin{figure}[h]
\includegraphics[width = \columnwidth]{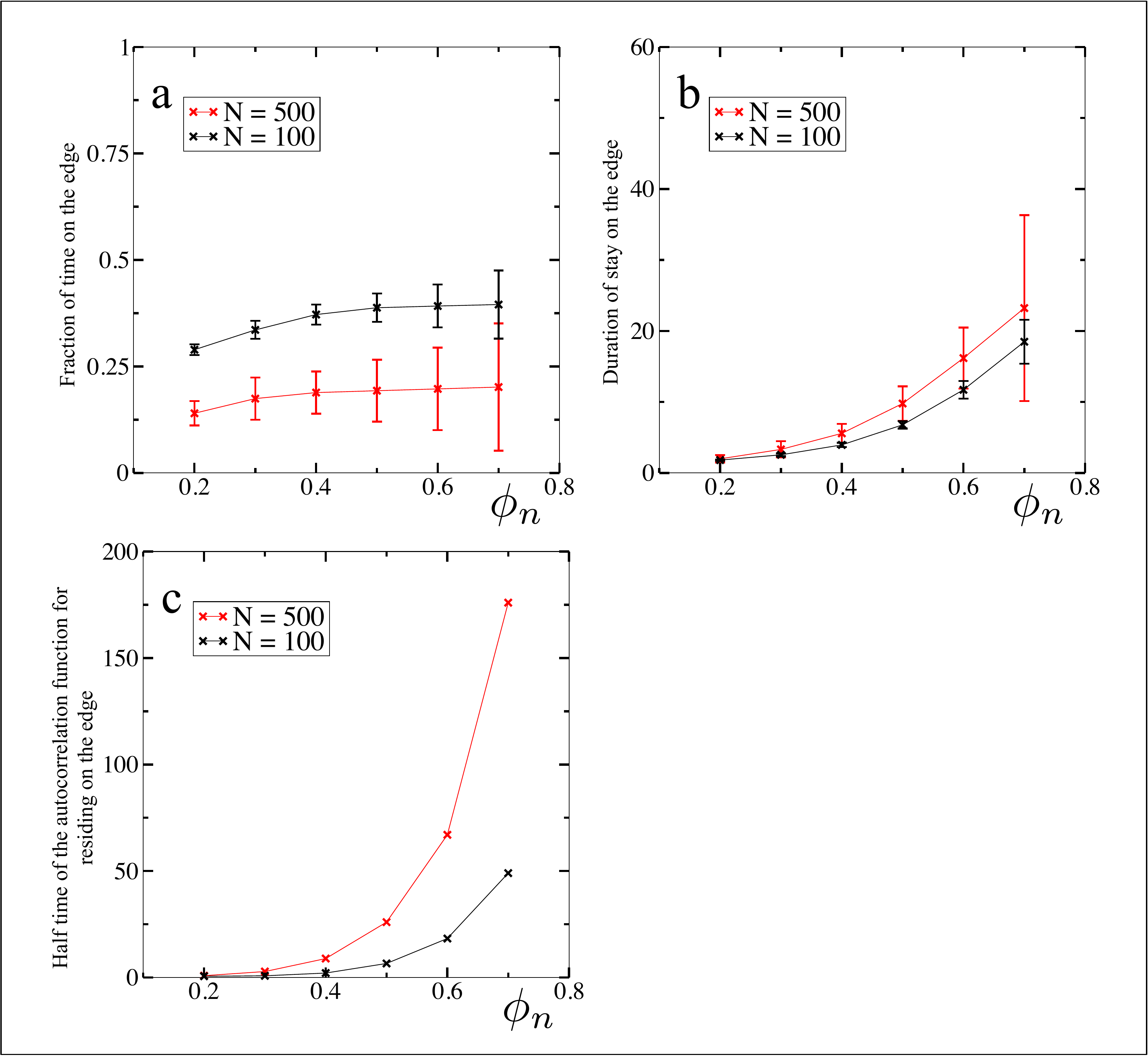}
\caption{\label{fig:edgetimes} a) Fraction of time a particle spends on the edge of the swarm b) average duration of each stay on the edge and c) half-time (in time steps) of the autocorrelation function for a particle residing on the edge of a swarm of size $N=100$ (black) and $N=500$ (red) for different noise levels, $\phi_n$. Each point is the average over a 10,000 time step simulation following a 10,000 time step equilibration period.}
\end{figure}

The fraction of time a particle within a given swarm spends on the edge is dependent on the size and noise level of the swarm is shown in Fig.~\ref{fig:edgetimes}a. As the noise is increased the amount of time that a particle spends on the edge increases slightly and each time a particle visits the edge it stays longer, Fig.~\ref{fig:edgetimes}b,c. One factor that contributes to this is that, as the noise is increased, the swarm takes on a more isotropic and diffuse structure by expanding into a roughly spherical shape. This higher surface area to volume ratio contributes to the increase in the fraction of particles on the edge, Fig.~\ref{fig:edgetimes}a.
The more spherical shape of these swarms also means the metric free surface term, $\langle\hat{\underline{r}} ^{t}_{ij} \rangle_{j \in S_i}$, is smaller because the inward-pointing component of the vectors connecting particles on the surface becomes smaller on average, see Fig~1 (main text). The reduction in density, consistent with Fig~1b (main text), means that a particle on the edge must travel further to pass another particle and re-join the bulk, hence this takes longer. The higher noise also means the effects of the metric free surface term are reduced and a particle on the edge is advected inwards more slowly. This leads to longer edge residence times.

\section{Role of parameters $\phi_e$ and $\phi_n$}

\begin{figure}[hB]
\includegraphics[width = \columnwidth]{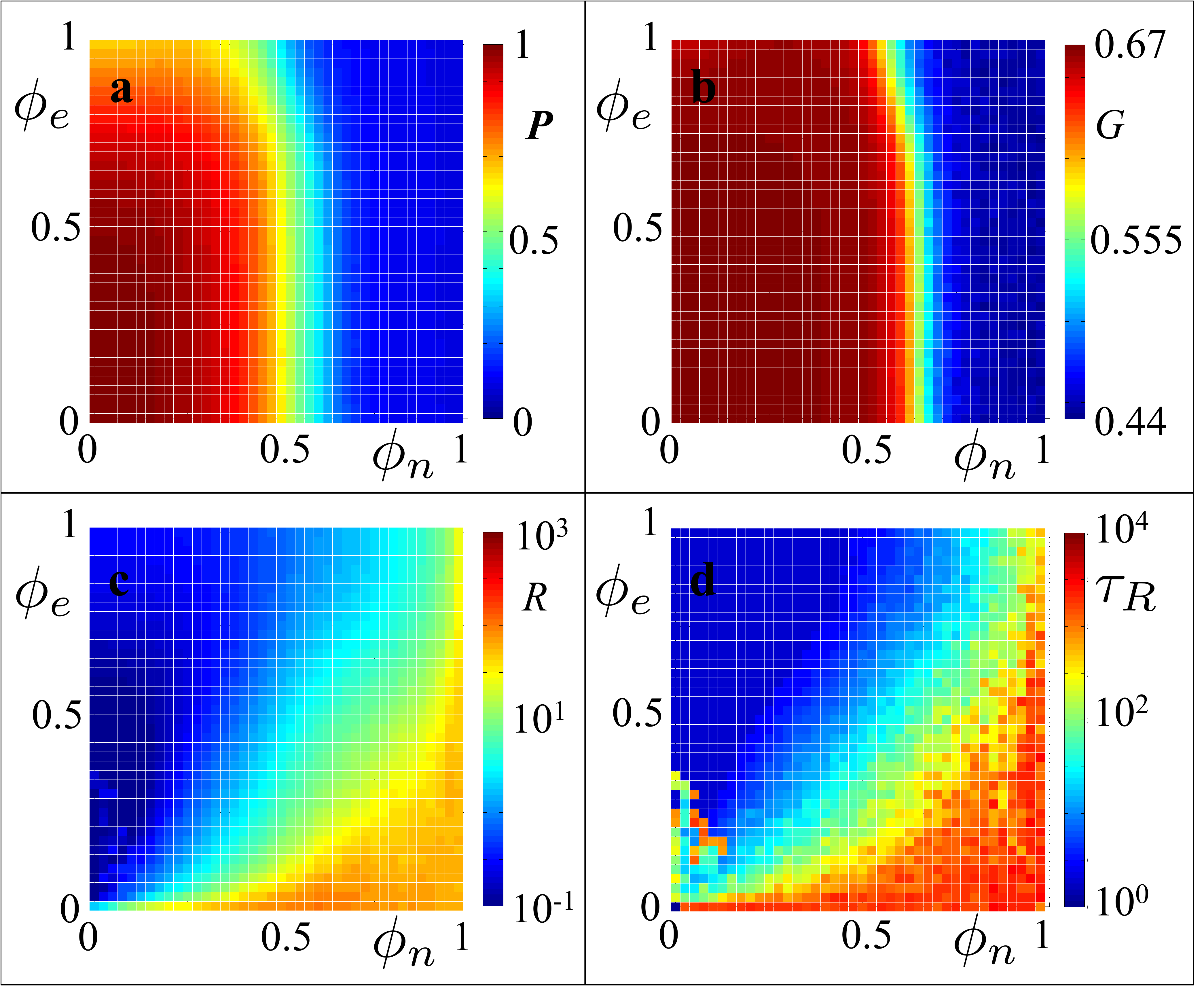}
\caption{\label{fig:phaseplanes} Variation of a) global order parameter $P$, b) binder cumulant $G$, c) average nearest neighbour separation, $R_1$ and d) half time of the autocorrelation function for $R_1$, $\tau_{R_1}$ on $\phi_e$ and $\phi_n$. Each point on the figure corresponds to an average quantity taken over 3 independent simulations of $N=100$ individuals over 15 000 time steps. }
\end{figure}

In order to asses how the properties of the flocks generated by our SMF model depend on the control parameters we simulate swarms at equilibrium for a full range of values for $\phi_e$ and $\phi_n$, controlling the strengths of the ``edge" and ``noise" terms, respectively. Each point on Fig.~\ref{fig:phaseplanes}a-d corresponds to a different pair of values for $\phi_e$ and $\phi_n$ for which we create 3 independent simulations each \hbox{15 000} time steps long, following a pre-equilibration phase of \hbox{15 000} time steps (which is higher than the nearest neighbour separation autocorrelation times shown in Fig.~\ref{fig:phaseplanes}d). 

The SMF model exhibits a transition from a disordered to an ordered phase around $\phi_n \approx 0.5$; see Fig.~\ref{fig:phaseplanes}a. The introduction of the metric free surface term, for which  $\phi_e \ne 0$, has very little effect on the existence of this transition. Through this transition, the binder cumulant varies monotonically from 1/3 to 2/3 indicating a continuous transition for all values of $\phi_e$, see Fig.~\ref{fig:phaseplanes}b. 
The position and sharpness of this transition change slightly with $\phi_e$: higher values of $\phi_e$ giving rise to a smoother order-disorder transition, occurring at a lower value of noise. However, there is no order-disorder transition associated with variation of $\phi_e$ alone. Indeed, for low noise, there is a high level of order even when $\phi_e=1$. This shows that the metric free surface term  does not have a significant effect on the ability of the swarm to achieve global order. This is because a high proportion of the swarm are not in the convex hull, hence will have $f_i = 0$; for a swarm of 100, this corresponds to between 60 and 80 particles, depending on $\phi_e$ and $\phi_n$.

All simulations featuring a significant surface term $(\phi_e > 0.1)$ have a well-controlled swarm density, indicated by the nearest neighbour separation, with low autocorrelation times Fig.~\ref{fig:phaseplanes}c,d; this indicates no long-term trends or periodicity. In the presence of high noise (disordered phase) we see the density decrease and density autocorrelation time increase as the surface term is reduced. In the ordered phase, the difference is less pronounced as reduced noise leads to a reduced rate of expansion. If the inward force is switched off entirely, $\phi_e = 0$, then the swarm is still able to achieve a high level of global order but the absence of any binding force means the density drops monotonically in the presence of any noise. Therefore there is no well defined density or density auto-correlation time for $\phi_e = 0$, see Fig.~\ref{fig:phaseplanes}c,d. When $\phi_e$ and $\phi_n$ are both low, the density and density autocorrelation times both show an area of increase. This is caused by ergodicity breaking in which the swarm occupies long-lived configurations that it is unable to escape from. For example, when co-alignment, and therefore global order, is very high the velocity of the centre of mass of the swarm must be very similar to the velocity of an individual. When this is the case, it becomes increasingly difficult for any individual to move relative to the centre of mass of the swarm as all the individuals' velocities are nearly parallel. Therefore it becomes less likely that the swarm will change shape. We call this a {\it hyper ordered state}.
\begin{figure}[h]
\includegraphics[width = \columnwidth]{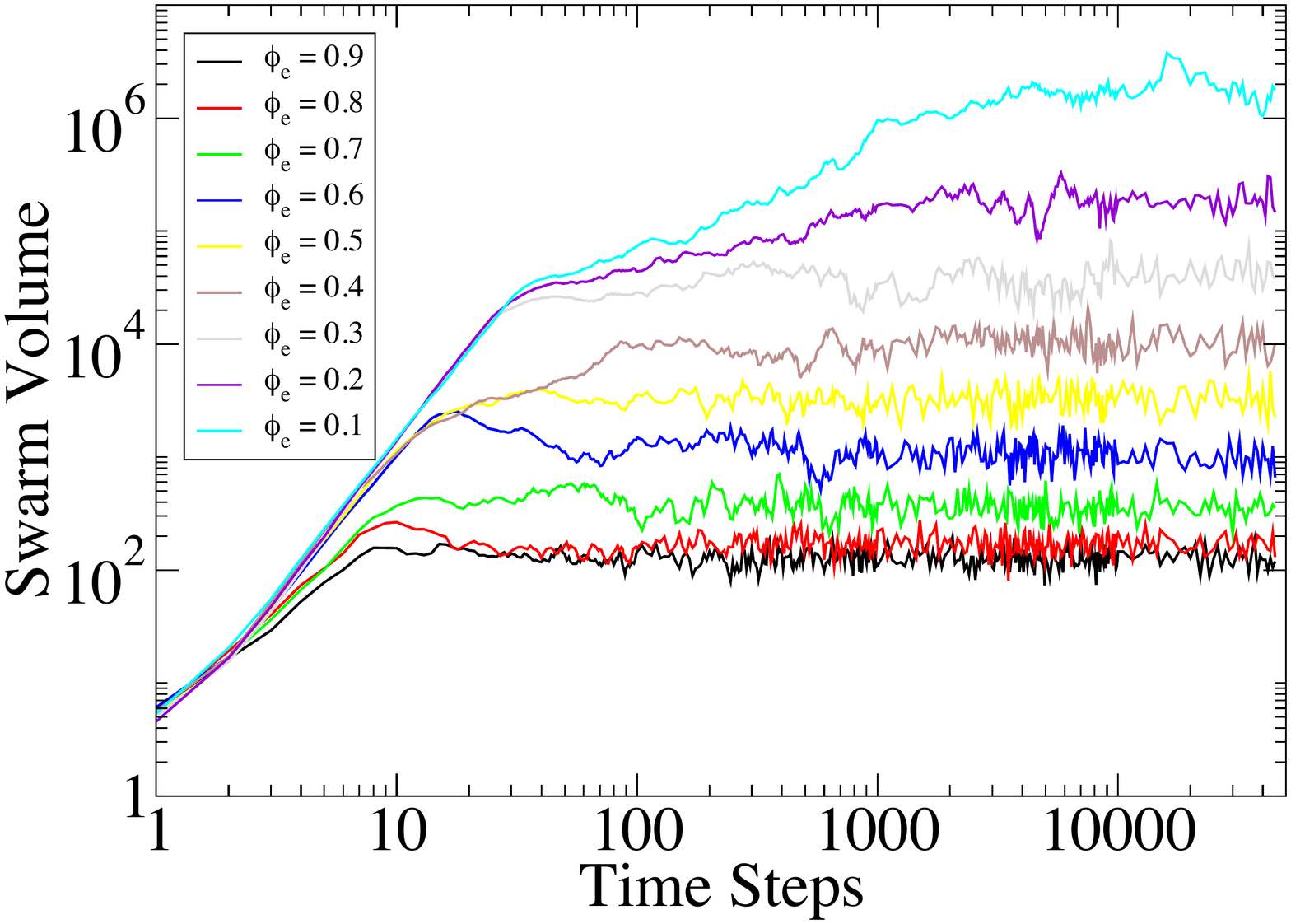}
\caption{\label{fig:relaxV} Volume of the convex hull of the swarm over the course of 30,000 time steps during pre-equilibration for varying levels of the surface term $(\phi_e \in (0.1,0.9),\phi_n =0.5, N=100)$. Relaxation is slow for the smallest values of $\phi_e$.}
\end{figure}
\section{Swarm equilibration}
\begin{figure}[h]
\includegraphics[width = \columnwidth]{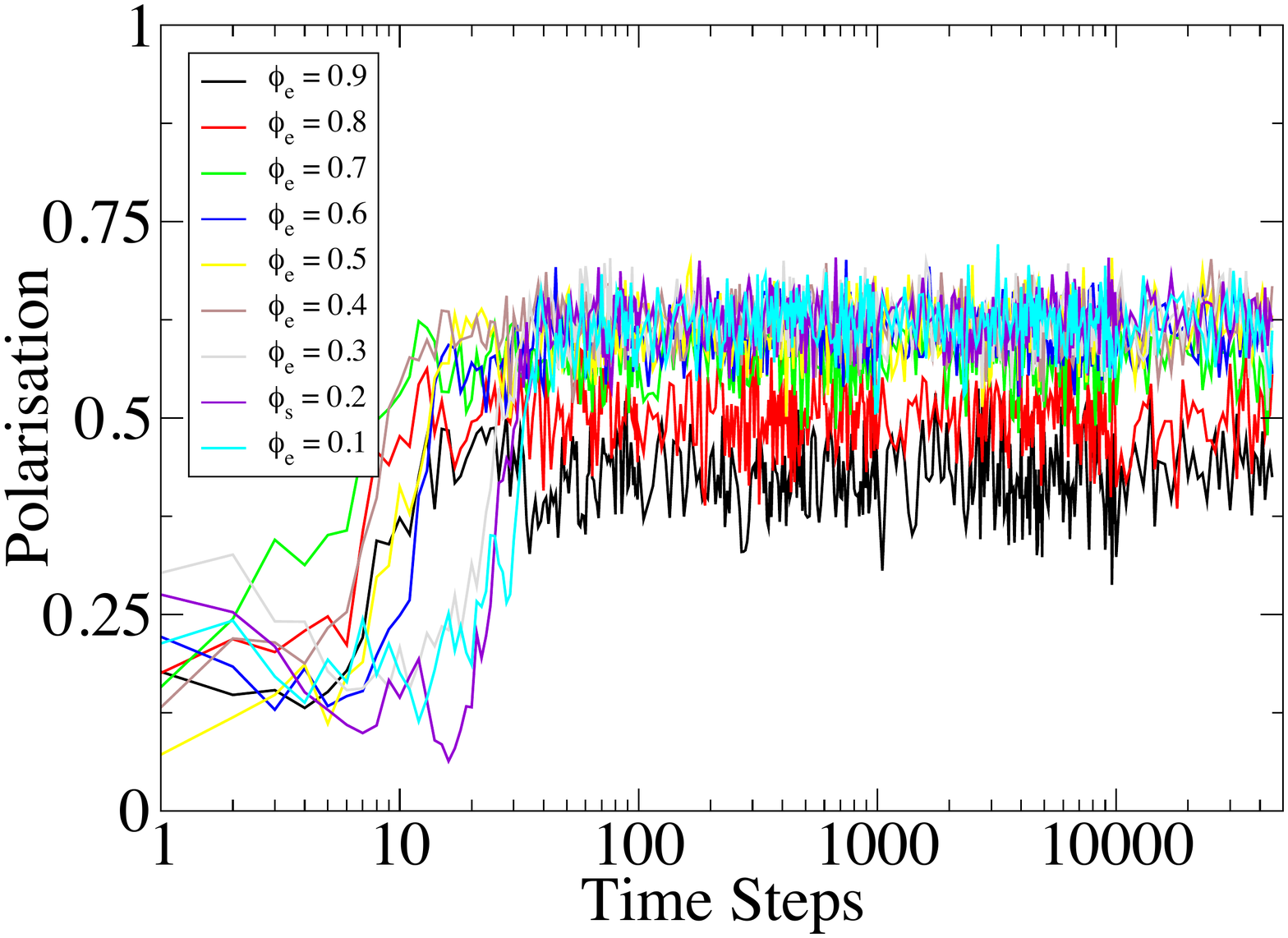}
\caption{\label{fig:relaxP} Polarisation, $P$, of the swarm over the course of 30,000 time steps during pre-equilibration for varying levels of the surface term $(\phi_e \in (0.1,0.9), \phi_n =0.5, N=100)$. Relaxation is faster than in Fig.~\ref{fig:relaxV}.}
\end{figure}

Due to the metric free nature of the model, there are two different relaxation mechanisms. The swarm expands in space until it reaches its steady state volume. 
This volume depends on the values for the parameters in the model, primarily $\phi_e$,$\phi_n$ and $N$. Swarms in which the effect of the surface confinement is reduced relative to the other terms give rise to higher volumes. Similarly swarms containing more individuals are larger. See Fig.~\ref{fig:relaxV}. 
Since the interactions between the particles are entirely topological the alignment interactions does not depend on the absolute distance between two particles and so we observe a significantly faster relaxation time for the ordering of the swarm, see Fig.~\ref{fig:relaxP}. 

Since the spatial relaxation time can become much longer than the orientational relaxation time we must pre-equilbrate our simulations for much longer periods than would be required for systems that lack a confining surface term. Such systems expand indefinitely and so only the shorter, orientational relaxation time is usually considered relevant.  This sets a correspondingly smaller limit on the swam size $N$ that is accessible to us, although we can still equilibrate swarms containing several thousand individuals (see main text).

\section{Supplementary Movies}
There are 4 SI movies provided. The SI movies 1 and 2 show typical swarms with $\phi_n = 0.5$ and $N=100$ and $N=500$, respectively. SI movies 3 and 4 show a swarm going through the order transition and back for $N=100$ and $N=500$, respectively (see captions for details).

\begin{figure}[h]
\includegraphics[width = \columnwidth]{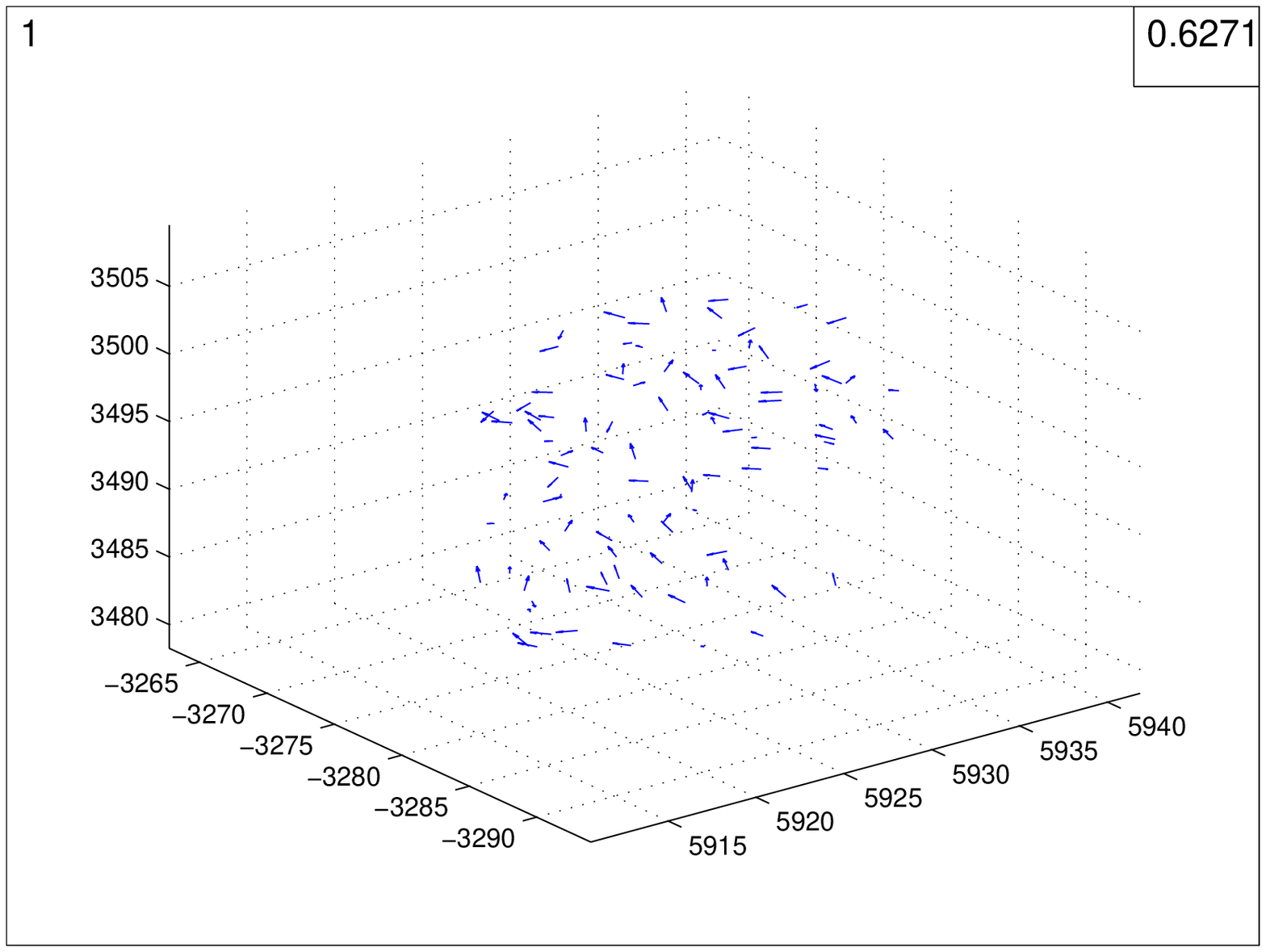}
\caption{\label{fig:SImov1} SI Movie 1 shows a swarm of 100 particles with $\phi_n = 0.5$ resulting in a polarisation around $P = 0.6$. The number in the top left shows the time step, and the polarisation is shown in the top right. 
}
\end{figure}
\begin{figure}[h]
\includegraphics[width = \columnwidth]{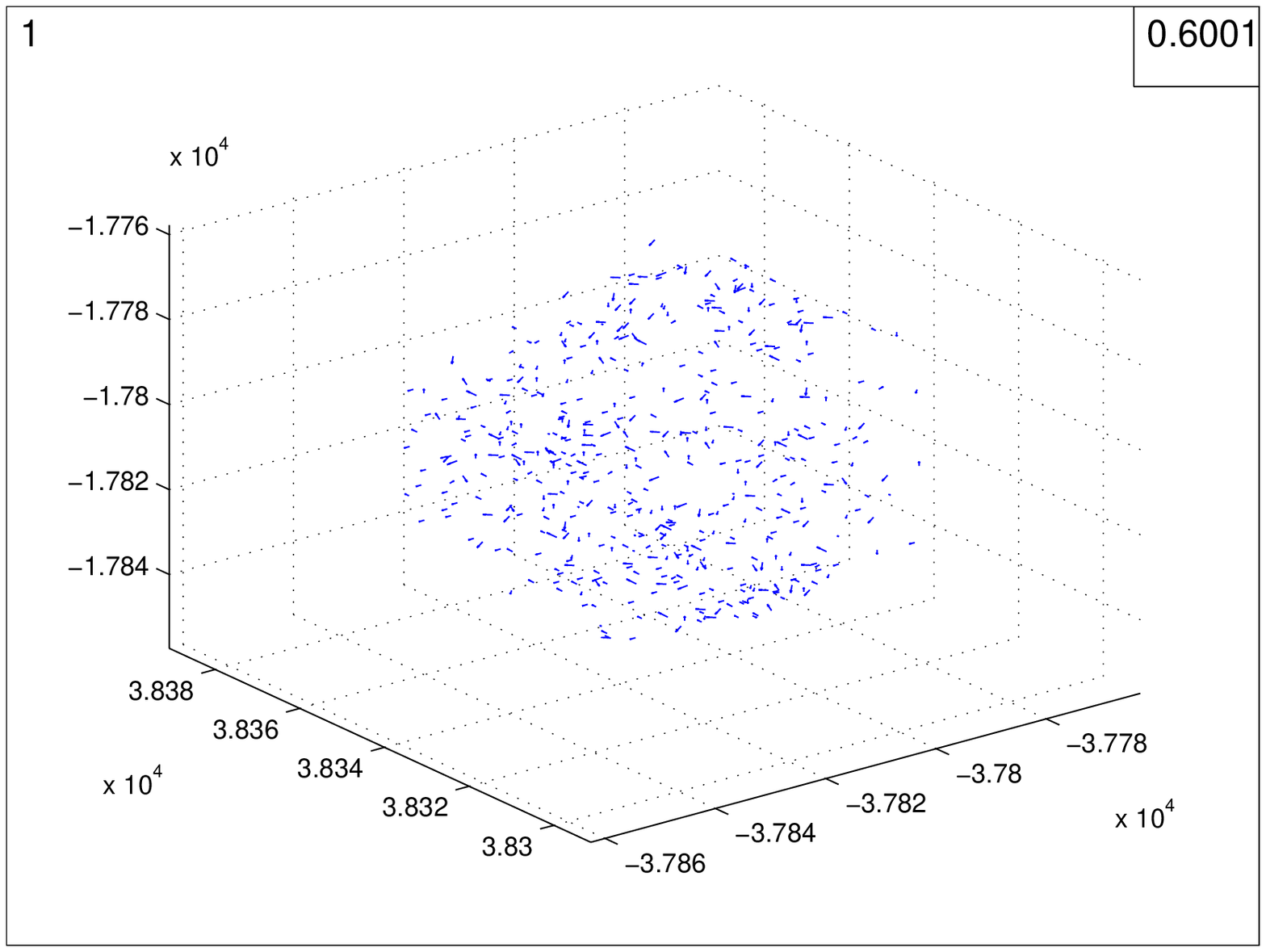}
\caption{\label{fig:SImov1} SI Movie 2 shows a swarm of 500 particles with $\phi_n = 0.5$ resulting in a polarisation around $P = 0.6$. The number in the top left shows the time step, and the polarisation is shown in the top right. 
}
\end{figure}
\begin{figure}[h]
\includegraphics[width = \columnwidth]{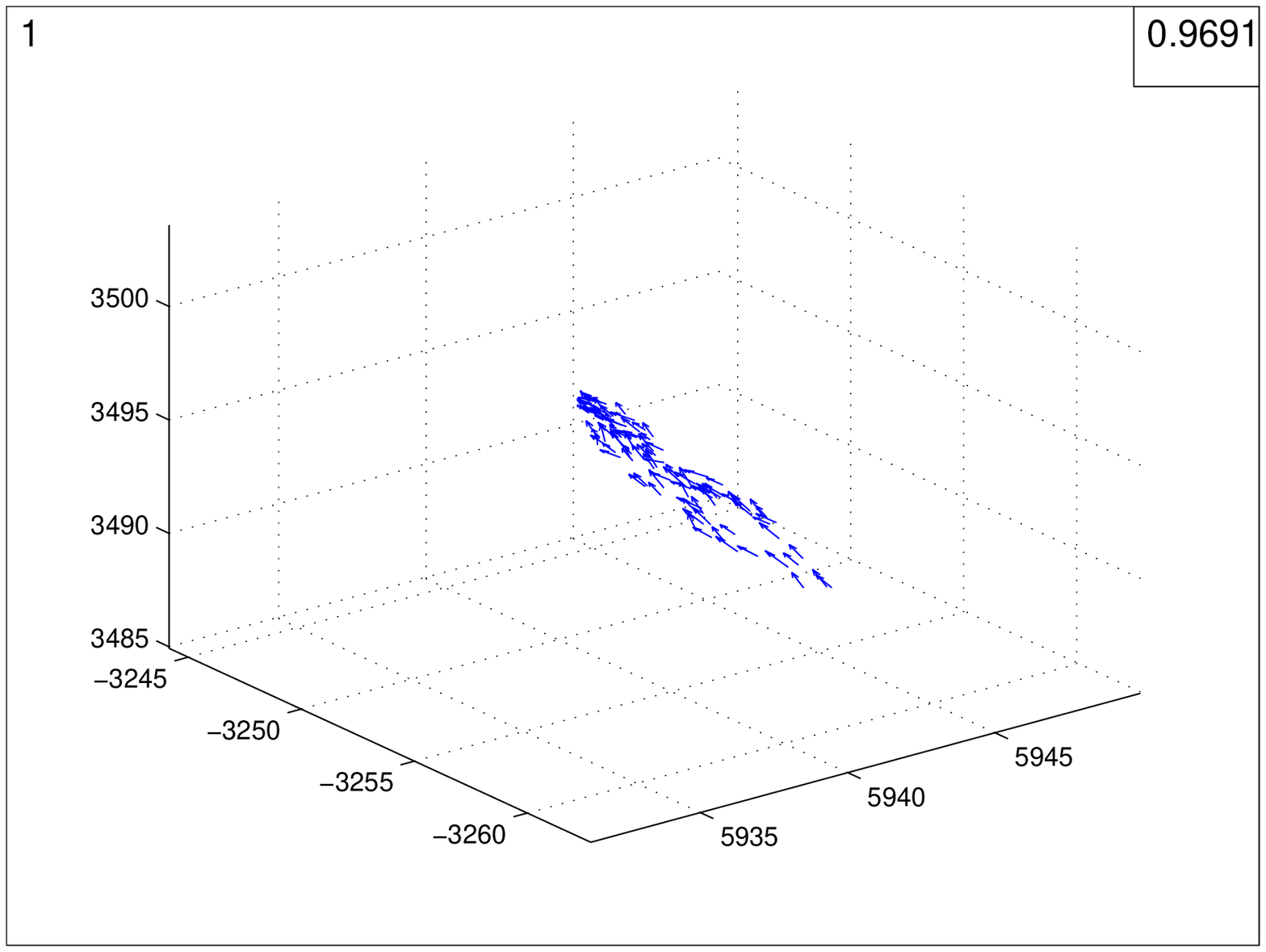}
\caption{\label{fig:SImov1} SI Movie 3 shows a swarm of 100 particles starting with $\phi_n = 0.15$, increasing slowly until $\phi_n = 0.85$ at the middle of the movie and then decreasing back to $\phi_n = 0.15$ for the end of the movie. This shows the behaviour of the swarm through the order transition. 
}
\end{figure}
\begin{figure}[h]
\includegraphics[width = \columnwidth]{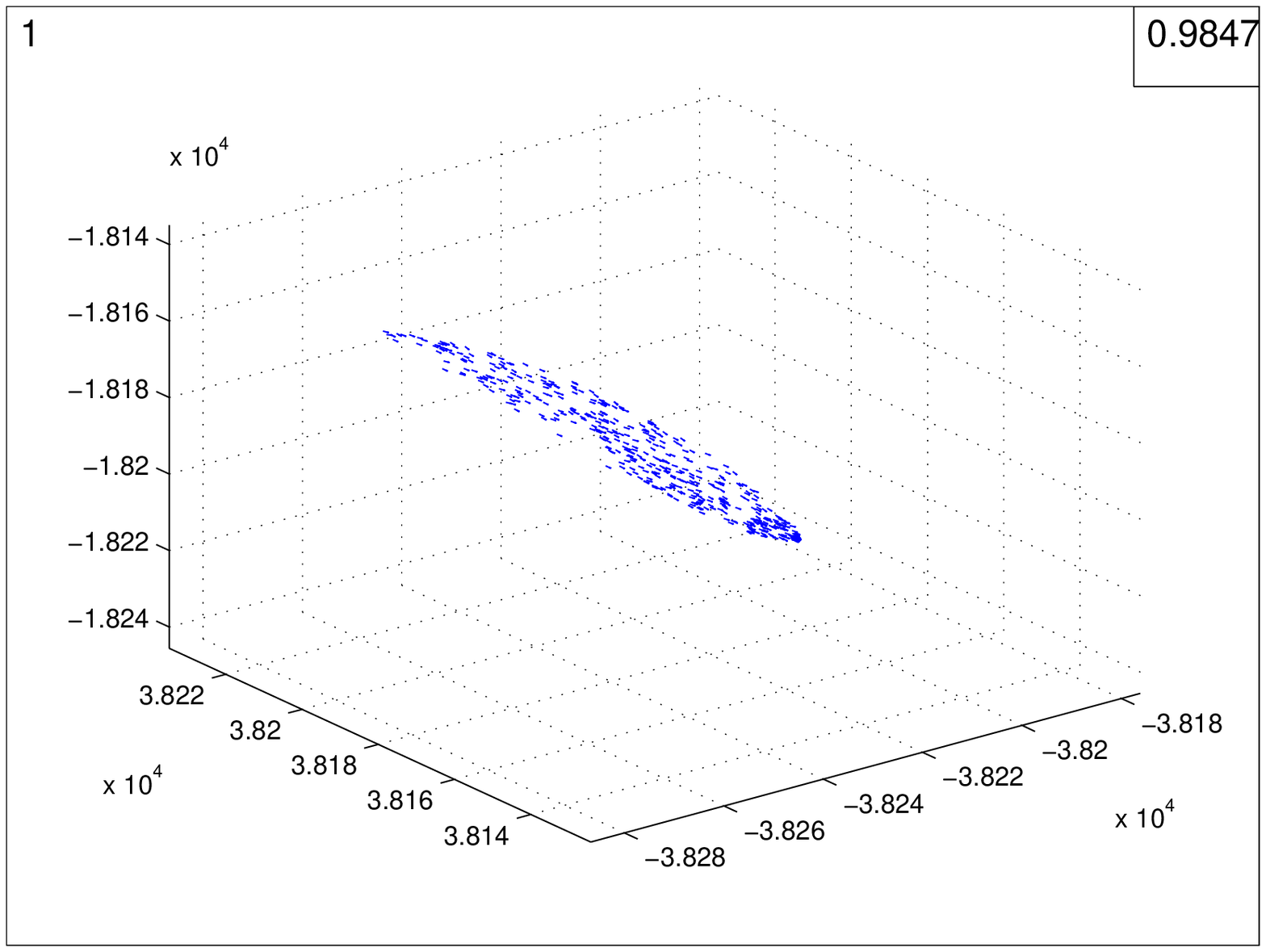}
\caption{\label{fig:SImov1} SI Movie 4 shows a swarm of 500 particles starting with $\phi_n = 0.15$, increasing slowly until $\phi_n = 0.85$ at the middle of the movie and then decreasing back to $\phi_n = 0.15$ for the end of the movie. This shows the behaviour of the swarm through the order transition. 
}
\end{figure}